\documentclass{article}

     \usepackage[nonatbib,preprint]{neurips_2020}

\usepackage[utf8]{inputenc} 
\usepackage[T1]{fontenc}    
\usepackage{hyperref}       
\usepackage{url}            
\usepackage{booktabs}       
\usepackage{amsfonts}       
\usepackage{nicefrac}       
\usepackage{microtype}      
\usepackage{graphicx}       
\usepackage{subcaption}     
\usepackage{adjustbox}
\usepackage{multirow, makecell}

\title{OutlierNets: Highly Compact Deep Autoencoder Network Architectures for On-Device Acoustic Anomaly Detection}

\author{Saad Abbasi$^{1}$, Mahmoud Famouri$^{2}$, Mohammad Javad Shafiee$^{1,2,3}$, and Alexander Wong$^{1,2,3}$\\
$^1$Department of Systems Design Engineering, University of Waterloo, Canada\\
$^2$Waterloo Artificial Intelligence Institute, Canada\\
$^3$DarwinAI Corp., Canada\\
}

\begin{document}

\maketitle

\begin{abstract}
Human operators often diagnose industrial machinery via anomalous sounds. Automated acoustic anomaly detection can lead to reliable maintenance of machinery. However, deep learning-driven anomaly detection methods often require an extensive amount of computational resources which prohibits their deployment in factories. Here we explore a machine-driven design exploration strategy to create \textbf{OutlierNets}, a family of highly compact deep convolutional autoencoder network architectures featuring as few as \textbf{686} parameters, model sizes as small as \textbf{2.7 KB}, and as low as \textbf{2.8 million} FLOPs, with a detection accuracy matching or exceeding published architectures with as many as 4 million parameters. Furthermore, CPU-accelerated latency experiments show that the OutlierNet architectures can achieve as much as \textbf{21x} lower latency than published networks.

\end{abstract}

\section{Introduction}
\label{submission}
Acoustic anomalies are one of the primary ways through which malfunctioning machinery or industrial processes are monitored. However, this detection of abnormal sounds is typically done subjectively via human operators who need prior experience. This is an important problem as automatic detection of acoustic abnormalities can lead to more reliable predictive maintenance and potentially eliminate the need for manual monitoring. More importantly, interconnected intelligent monitoring systems are also a primary enabling technology for Industry 4.0 (AI-based factory automation).

A variety of deep learning-driven techniques have been introduced for acoustic anomaly detection (AAD) in recent years, including dense autoencoders~\cite{Purohit2019MIMIIInspection,Ribeiro2020DeepSounds}, convolutional autoencoders~\cite{Ribeiro2020DeepSounds}, and pre-trained convolutional neural networks~\cite{Muller2020AcousticLearning}. Although deep learning-driven methods have demonstrated excellent accuracy in detecting anomalous sounds, the widespread adoption of these methods remains limited. One of the primary reasons for the slow adoption is the prohibitive computational resources required by many deep learning-driven anomaly detection methods, which often have high architectural and computational complexities, as well as high memory footprints. This results in methods that cannot be deployed on resource-constrained edge computing devices such as CPUs or microcontrollers. Moreover, since such methods are typically designed without resource constraints, the on-device prediction latency is typically not an important design parameter. However, for industry deployment, AAD must exhibit real-time latency to be successful, as a quick shutdown of an abnormal sounding machine could be crucial for safety.

Motivated by this challenge, we explore a machine-driven design exploration strategy to create optimal macro-architecture and micro-architecture designs tailored specifically for on-device AAD tasks. Through this strategy, we introduce OutlierNets, a set of highly compact deep convolutional autoencoder network architectures that provide high architecture and computational efficiency while providing high unsupervised AAD accuracy.  In addition to assessing \textit{area under the receiver operating curve} (AUC) performance, we further assess the CPU-accelerated latency performance of the OutlierNet architectures by leveraging a domain-specific compiler in the form of the OpenVINO platform, as well latency performance as on an ARM Cortex A-72 embedded CPU.

\subsection{Related Work}
\label{sec:related_work}
Until recently, the area of deep learning-driven AAD has been relatively underrepresented in the research literature. This has primarily been due to the lack of publicly available datasets. However, this situation has improved significantly in recent years with the release of AAD datasets such as MIMII \cite{Purohit2019MIMIIInspection}, and ToyADMOS \cite{Koizumi2019ToyADMOS:Detection}. Traditionally, autoencoders have been the cornerstone for anomaly detection \cite{Ribeiro2020DeepSounds,Duman2020AcousticProcesses,Bayram2021RealAutoencoders,Meire2019ComparisonAssets}. In the context of AAD, autoencoders are typically trained in an unsupervised manner on the normal operating sounds of machines. The key idea here is that an autoencoder will learn to reconstruct a machine's normal sounds with little reconstruction error. If the autoencoder is fed with an unseen (anomalous) sound, the reconstruction error would be significantly higher, leading to the detection of anomalous events. As a result, the reconstruction error of anomalous sounds would have a different distribution than for normal or ambient sound.

Autoencoders are typically trained on the features extracted from raw audio signals, with spectral features such as Mel-frequency Cepstral Coefficients \cite{Ribeiro2020DeepSounds} and Mel-spectrograms \cite{Ribeiro2020DeepSounds, Duman2020AcousticProcesses, Bayram2021RealAutoencoders} amongst the most popular. A Mel-spectogram is similar to a conventional spectogram, with the major difference being that the sound is represented on the Mel-scale, which measures the pitch as perceived by humans. The advantage offered by a Mel-spectrogram is that it transforms the time-series representation of the acoustic signal into a rich higher-dimensional representation of frequency, time, and intensity (power), which makes them well-suited for convolutional autoencoder architectures.

Muller~\textit{et al.} ~\cite{Muller2020AcousticLearning} proposed the use of ImageNet pre-trained CNNs (e.g., ResNet-18 \cite{He2016} and AlexNet \cite{krizhevsky2012imagenet}) for automatic feature extraction from Mel-spectrograms. The extracted features are subsequently fed to traditional machine learning methods such as Gaussian Mixture Models and Support Vector Machines for inference. Although the aforementioned deep learning-driven methods have demonstrated excellent accuracy, the resulting systems require a large amount of computational and memory resources to operate. For example, a ResNet-18 based feature extractor would require 1.8 billion FLOPs for inference and exhibits a memory footprint of over 500 MB.  To address the high complexity of such pre-trained CNN architectures, Ribeiro \textit{et al.} \cite{Ribeiro2020DeepSounds} proposed a tailored deep convolutional autonencoder architecture for AAD with a much lower architectural complexity at 4M parameters and a memory footprint of 15 MB.  Banbury \textit{et al.} \cite{Banbury2021Micronets:Microcontrollers} leveraged a differentiable neural architecture search strategy~\cite{liu2019darts} to design neural network architectures for microcontrollers, the smallest of which is a dense autoencoder architecture with an impressive size of 270 KB and the AUC of 84.7\%.
\section{Methods}
\label{sec:methods}
In this study, we present OutlierNets, a family of highly compact deep convolutional autoencoder architectures tailored for real-time detection of acoustic anomalies. These networks were designed with a machine-driven design exploration strategy. Human expertise was leveraged to design  operational constraints and initial design prototype while the machine-driven design exploration process traverses the architectural search space for optimal network design.

\subsection{Dataset}
To evaluate the OutlierNet designs, we employ the MIMII dataset \cite{Purohit2019MIMIIInspection}.
The dataset consists of normal and malfunctioning sounds of industrial fans, valves, sliders, and pumps. In this study, we focus on the slider and fan machine types. Each machine type is comprised of recordings for four different machines. For example, for the slider machine type, recordings of four distinct slider machines are provided. The training set comprises normal recordings exclusively, whereas the test set is an even mix of malfunctioning and normal sounds.  To quantify our detection accuracy, we use the commonly used area under the receiver operating curve (AUC) metric.

\subsection{Mel-spectrograms}
Similar to \cite{Ribeiro2020DeepSounds, Bayram2021RealAutoencoders}, we train OutlierNets on Mel-spectrograms of acoustic recordings rather than time-series representations. We employ 128 Mel bands with a hop length of 512, with the Fourier window set to 1024, resulting in $313 \times 128$ Mel-spectrograms. The resulting Mel-spectrogram is then cropped into $32 \times 128$ windows with no overlap. Each window represents approximately one second of audio. This is crucial for two reasons. First, the shorter window provides a more uniform image for an autoencoder to learn. Second, a shorter window demonstrates that our system requires only one second of recorded audio to detect anomalous sound. This is an important consideration for real-time AAD. A system that requires a considerably longer recording would have an inherent lag built into the system and would not be real-time, even if the prediction itself is rapid. An example of a Mel-spectrogram is given in Figure~\ref{fig:autoencoder} along with the cropping boxes. Note that the small window would not deteriorate the autoencoder's ability to detect anomalous sound, even if the duration of the acoustic anomaly is longer than one second. This is because if an acoustic anomaly completely encompasses a 32 $\times$ 128 window, the autoencoder would still exhibit a substantially high reconstruction error, assuming that the anomalous sound is different from the normal operating sounds of the machine.

\subsection{Machine-driven Design Exploration}
Prior to advances in machine-driven exploration of architectural search spaces, the design of neural network architectures was a very challenging and time-consuming process. The additional resource constraints necessary for real-time edge scenarios (e.g., architectural and computational complexities, memory footprint, etc.) makes this process even more challenging.  Given the number of design hyperparameters involved, the architectural search space of even simple neural networks is often extremely large. This led to great recent interest in machine-driven methods for exploring the architectural search space to find the optimal network given a set of constraints or design objectives.

The problem of traversing the architectural search space for an optimal neural network can be generally framed in two different ways. The first category of approaches~\cite{Tan2018MnasNet:Mobile,Hsu2018MONAS:Learning, Banbury2021Micronets:Microcontrollers, elsken2019efficient} formulates the search as a multi-objective optimization problem. The objectives can include criteria such as network size, inference latency, accuracy or FLOPs. In these approaches, optimization is performed via reinforcement learning, evolutionary algorithms, and gradient descent. The second category of approaches~\cite{Wong2018FermiNets:Synthesis} leverages a constrained optimization problem formulation, which consists of an objective function~(e.g., \cite{wong2019netscore}) as well as a set of operational constraints such as accuracy requirements and computational and architectural requirements.

In this investigation, we leverage the second approach via generative synthesis \cite{Wong2018FermiNets:Synthesis} to explore the architecture design space and obtain optimal network designs for on-device real-time AAD. The ultimate goal of this study is to achieve an optimal balance between accuracy and efficiency. To achieve this, we leveraged the performance function proposed in~\cite{wong2019netscore} which takes into account the tradeoff between accuracy, computational complexity, and architectural complexity, and further imposed network parameter count constraints ($<100,000$ parameters) and AUC constraints (within 10\% of state-of-the-art convolutional autoencoder architecture in~\cite{Ribeiro2020DeepSounds}, which has 4M parameters).  More specifically, generative synthesis can be formulated as a constrained optimization problem, where the goal is to obtain a tailored deep neural network architecture as determined by an optimal generator $\mathcal{G}$ which, given seeds $\mathcal{S}$, generates neural network architectures that maximize a performance function $\mathcal{U}$ while satisfying the constraints defined in an indicator function $1_r(\cdot)$,
\begin{equation}
\mathcal{G}=\max_{\mathcal{G}}\mathcal{U}(\mathcal{G}(s)) \textrm{ subject to $1_r(G(s))=1, \forall \in \mathcal{S}$}.
\end{equation}
Given an initial design prototype $\varphi$, $\mathcal{U}$, and $1_r(\cdot{})$, the approximate solution to this constrained optimization problem can be found in an iterative fashion. In this study, we formulate $1_r(\cdot{})$ to impose the parameter count and AUC constraints mentioned earlier. For the initial design prototype $\varphi$, we leverage human experience and expertise and we define a convolutional autoencoder design prototype whose input is a $32 \times 128$ Mel-spectrogram, with convolutional kernel sizes defined as $3 \times 3$.

\begin{figure}[t]
\centering\includegraphics[width=\textwidth]{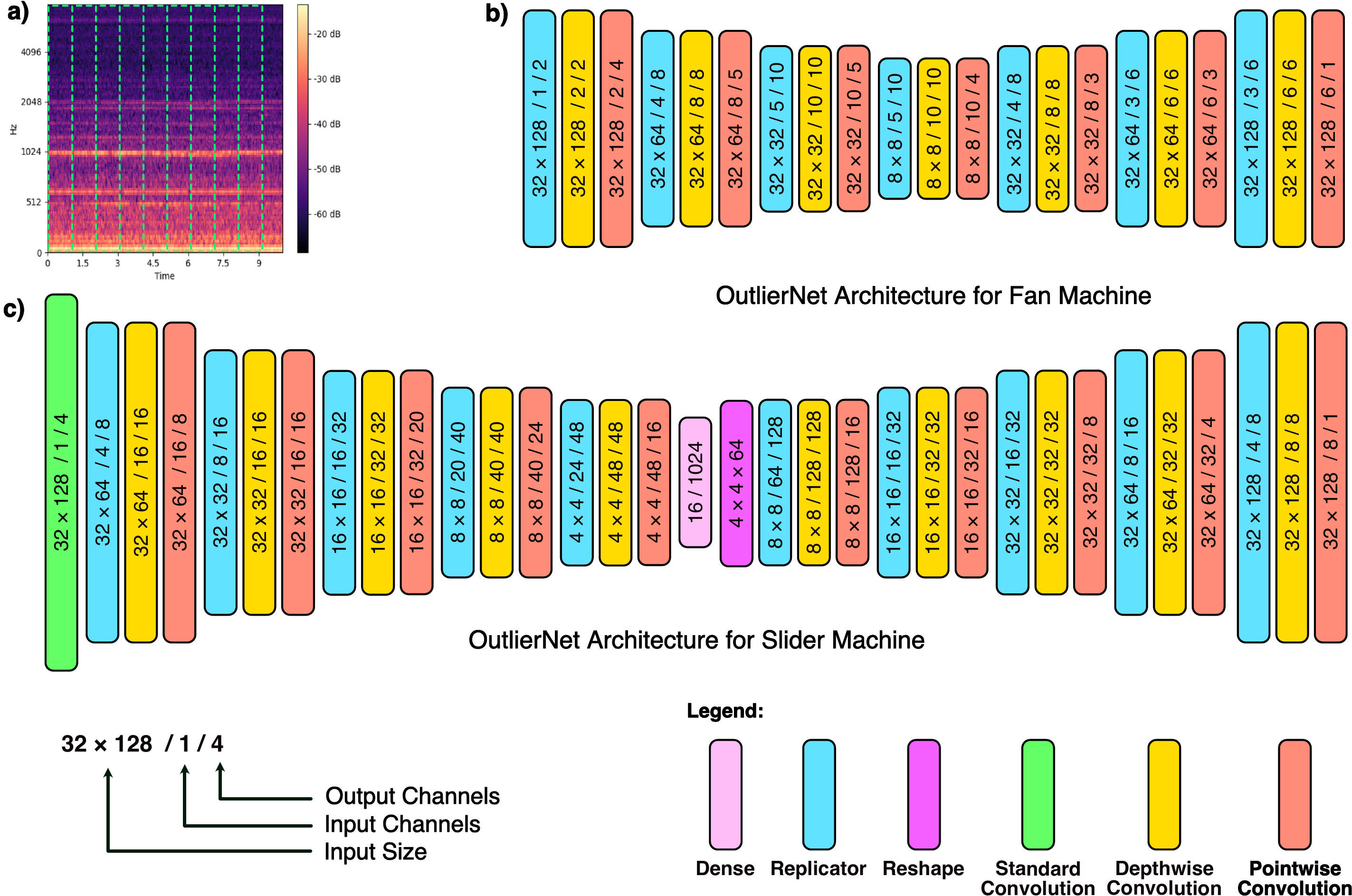}
\caption{a) Mel-spectrogram of anomalous fan sound. Green dashed boxes represent $32 \times 128$ crops, each representing approximately one second of audio. b) Example of an OutlierNet autoencoder architecture for fan AAD tasks (ID: 06, SNR: 6 dB) c) Example of an OutlierNet autoencoder architecture for slider AAD tasks (ID: 00, SNR: -6 dB).}
\label{fig:autoencoder}
\end{figure}

Examples of the final OutlierNet architecture designs for AAD for fans and for sliders are shown in Figure ~\ref{fig:autoencoder}b and \ref{fig:autoencoder}c. A number of observations can be made about the OutlierNet architecture designs.  First, the OutlierNet designs possess very lightweight macro-architecture designs comprised largely of depthwise convolutions, pointwise convolutions, and replicators for high efficiency.  Second, the OutlierNet designs exhibit high micro-architecture design diversity to further reinforce a strong balance between AAD accuracy and efficiency.  Third, the OutlierNet designs exhibit macro-architecture differences between designs tailored for fan AAD tasks and designs for slider AAD tasks, with architecture designs for slider ADD tasks being more complex (e.g., additional standard convolution layers and a dense latent space) given that slider ADD is a tougher task. The OutlierNet network architectures are then passed through a domain-specific compiler built for Intel CPUs (i.e., OpenVINO) for CPU-accelerated latency experiments. This compiler yields CPU-accelerated models that feature a significantly lower inference latency and demonstrate efficiency of the proposed OutlierNet architectures. Furthermore, the architectures were evaluated on an ARM Cortex A-72 embedded CPU to assess their efficacy on low-power, resource-constrained environments.

\section{Results and Discussion}
To evaluate the efficacy of the OutlierNet architecture designs, we compare their AUC, parameter counts, model sizes, FLOPs, and on-device latencies with the state-of-the-art deep convolutional autoencoder architecture introduced in~\cite{Ribeiro2020DeepSounds}. Similar to our approach,~\cite{Ribeiro2020DeepSounds} employ small overlapping Mel-spectrograms to classify anomalous sounds. For brevity, we refer to their proposed architecture as \textbf{CAE-MCS} as a shorthand for convolutional autoencoder for machine condition sounds for the remainder of this study. To provide a fair comparison, we implement and evaluate CAE-MCS on the fan and slider datasets from the MIMII dataset. We found this architecture to achieve high AAD performance, reporting an average AUC of 89.1\% for slider AAD tasks and 83.5\% for fan AAD tasks while only having 4M parameters. The results of OutlierNets and CAE-MCS are compared in Table~\ref{tab:results}.

\hspace{-0.2in}
\begin{table*}[h!]
\caption{Performance comparison between OutlierNet architectures with the convolutional autoencoder architecture in \cite{Ribeiro2020DeepSounds} in terms of AUC, architectural complexity, computational complexity, and on-device latency with ARM Cortex-A72 embedded CPU as well as an Intel Core i5-7600K desktop CPU. These measurements are included in the columns \textit{ARM ($\mu$s)} and \textit{Intel ($\mu$s)}. Top half of the table provides a comparison with the Fan machine type whereas the bottom half evaluates the architectures on the Slider machine type.}
\begin{center}
\begin{small}
\begin{sc}
\vspace{-0.15in}
\begin{tabular}{rp{0.25cm}|rrlrrr|rr}
\toprule
      &    & \multicolumn{1}{l}{}       & \multicolumn{1}{l}{}                                                    & \multicolumn{3}{c}{\textbf{OutlierNets}}                                                                                             & \multicolumn{1}{l|}{}      & \multicolumn{2}{c}{\textbf{CAE-MCS }}                                                                                                         \\
SNR   & ID & \multicolumn{1}{l}{Params} & \multicolumn{1}{l}{\begin{tabular}[c]{@{}l@{}}Size \\(KB)\end{tabular}} & FLOPs & \begin{tabular}[c]{@{}l@{}}ARM\\($\mu$s)\end{tabular} & \multicolumn{1}{l}{\begin{tabular}[c]{@{}l@{}}Intel \\($\mu$s)\end{tabular}} & \multicolumn{1}{l|}{AUC}   & \multicolumn{1}{l}{AUC}             & \multicolumn{1}{l}{}                                                                                    \\
\hline\hline
\multicolumn{10}{c}{\textbf{Fan Machine Type}}                                                                                                                                                                                                                                                                                                                                                                                                 \\
\hline
6 dB  & 00 & 2,028                      & 7.9                                                                     & 7.33M & 11.62                                             & 0.530                                                                    & \textbf{87.0\%}            & 83.8\%                              & \multirow{12}{*}{\begin{tabular}[c]{@{}r@{}}\scriptsize{4M params}\\ \scriptsize{7.7 $\mu$s Intel} \\ \scriptsize{78.2 $\mu$s ARM}\\ \scriptsize{15MB Model Size}\end{tabular}}  \\
6 dB  & 02 & 907                        & 3.5                                                                     & 3.63M & 7.156                                             & 0.379                                                                    & 98.7\%                     & 98.7\%                              &                                                                                                         \\
6 dB  & 04 & 3,055                      & 11.9                                                                    & 10.6M & 12.445                                            & 0.623                                                                    & \textbf{96.6\%}            & 93.8\%                              &                                                                                                         \\
6 dB  & 06 & 686                        & 2.7                                                                     & 2.87M & 7.601                                             & 0.366                                                                    & \textbf{100.0\%}           & 99.8\%                              &                                                                                                         \\
0 dB  & 00 & 2,992                      & 11.7                                                                    & 10.3M & 18.149                                            & 0.570                                                                    & 62.2\%                     & \textbf{62.7\%}                     &                                                                                                         \\
0 dB  & 02 & 1,067                      & 4.2                                                                     & 4.18M & 12.305                                            & 0.414                                                                    & 91.9\%                     & \textbf{92.7\%}                     &                                                                                                         \\
0 dB  & 04 & 1,025                      & 4.0                                                                     & 4.19M & 12.783                                            & 0.426                                                                    & 79.7\%                     & \textbf{82.1\%}                     &                                                                                                         \\
0 dB  & 06 & 861                        & 3.4                                                                     & 3.39M & 12.847                                            & 0.382                                                                    & \textbf{99.6\%}            & 99.0\%                              &                                                                                                         \\
-6 dB & 00 & 6,840                      & 26.7                                                                    & 22.9M & 15.639                                            & 0.746                                                                    & 51.5\%                     & \textbf{54.5\%}                     &                                                                                                         \\
-6 dB & 02 & 2,892                      & 11.3                                                                    & 10.5M & 19.074                                            & 0.616                                                                    & 70.6\%                     & \textbf{77.4\%}                     &                                                                                                         \\
-6 dB & 04 & 1,826                      & 7.1                                                                     & 6.48M & 16.372                                            & 0.483                                                                    & 61.1\%                     & \textbf{63.9\%}                     &                                                                                                         \\
-6 dB & 06 & 2,919                      & 11.4                                                                    & 10.1M & 17.724                                            & 0.577                                                                    & \textbf{96.7\%}            & 93.4\%                              &                                                                                                         \\
\hline
      &    & \multicolumn{5}{r}{Average AUC}                                                                                                                                                                                                             & \multicolumn{1}{r}{83.0\%} & \multicolumn{1}{l}{\textbf{83.5\%}} & \multicolumn{1}{l}{}                                                                                    \\
\hline\hline
\multicolumn{10}{c}{\textbf{Slider Machine Type}}                                                                                                                                                                                                                                                                                                                                                                                              \\
\hline
6 dB  & 00 & 21,781                     & 85.1                                                                    & 4.63M & 10.513                                            & 0.481                                                                    & 99.4\%                     & \textbf{99.7\%}                     & \multirow{12}{*}{\begin{tabular}[c]{@{}r@{}} \scriptsize{4M params} \\\scriptsize{7.7 $\mu$s Intel} \\ \scriptsize{78.2 $\mu$s ARM}\\ \scriptsize{15MB Model Size}\end{tabular}}  \\
6 dB  & 02 & 69,877                     & 273.0                                                                   & 5.35M & 10.993                                            & 0.508                                                                    & 97.0\%                     & \textbf{98.6\%}                     &                                                                                                         \\
6 dB  & 04 & 58,729                     & 229.4                                                                   & 5.12M & 10.842                                            & 0.473                                                                    & \textbf{98.2\%}            & 97.7\%                              &                                                                                                         \\
6 dB  & 06 & 53,705                     & 209.8                                                                   & 5.11M & 10.893                                            & 0.501                                                                    & \textbf{91.4\%}            & 89.6\%                              &                                                                                                         \\
0 dB  & 00 & 21,781                     & 85.1                                                                    & 4.62M & 10.567                                            & 0.459                                                                    & 98.9\%                     & \textbf{99.3\%}                     &                                                                                                         \\
0 dB  & 02 & 69,877                     & 273.0                                                                   & 5.35M & 10.932                                            & 0.488                                                                    & 88.3\%                     & \textbf{92.8\%}                     &                                                                                                         \\
0 dB  & 04 & 42,377                     & 165.5                                                                   & 4.90M & 10.605                                            & 0.503                                                                    & \textbf{94.1\%}            & 93.1\%                              &                                                                                                         \\
0 dB  & 06 & 58,157                     & 227.2                                                                   & 5.17M & 10.758                                            & 0.498                                                                    & \textbf{75.0\%}            & 72.9\%                              &                                                                                                         \\
-6 dB & 00 & 27,229                     & 106.4                                                                   & 4.67M & 10.729                                            & 0.459                                                                    & \textbf{96.4\%}            & 95.8\%                              &                                                                                                         \\
-6 dB & 02 & 27,229                     & 106.4                                                                   & 4.67M & 10.729                                            & 0.463                                                                    & 78.5\%                     & \textbf{80.6\%}                     &                                                                                                         \\
-6 dB & 04 & 57,497                     & 224.6                                                                   & 5.11M & 10.748                                            & 0.473                                                                    & \textbf{86.9\%}            & 84.8\%                              &                                                                                                         \\
-6 dB & 06 & 58,157                     & 227.2                                                                   & 5.11m & 10.903                                            & 0.475                                                                    & 61.9\%                     & \textbf{63.7\%}                     &                                                                                                         \\
\hline
      &    & \multicolumn{5}{r}{Average AUC}                                                                                                                                                                                                             & \multicolumn{1}{r}{88.8\%} & \multicolumn{1}{l}{\textbf{89.1\%}} & \multicolumn{1}{l}{}                                                                                    \\
\bottomrule
\end{tabular}
\end{sc}
\end{small}
\end{center}
\label{tab:results}
\vspace{-0.2in}
\end{table*}

Several important observations can be made from these results. First of all, it can be observed that the proposed OutlierNet architectures achieved AUC scores matching or exceeding the much larger CAE-MCS deep convolutional autoencoder architecture for both the slider and the fan datasets despite being orders of magnitude lower in architectural and computational complexities. In particular, OutlierNets achieve \textbf{88.8\%} average AUC, across all slider machines and SNRs, whereas CAE-MCS achieves 89.1\%. Similarly, the proposed OutlierNet architectures achieve \textbf{83.0\%} average AUC for fans while CAE-MCS achieves 83.5\%. This demonstrates the effectiveness of a machine-driven design exploration strategy in constructing deep neural network architectures that strikes a strong balance between accuracy and efficiency.

Second, it can be observed that the proposed OutlierNet architectures exhibit very diverse architecture and computational complexities for each of the 24 AAD tasks tested in this study, all tailored around the complexity of the task at hand.  Specifically, CAE-MCS has a parameter count of approximately 4 million, whereas the smallest and largest OutlierNets in comparison have 686 parameters \textbf{(5800$\times$ fewer)} and 70,000 parameters \textbf{(57$\times$ fewer)}, respectively. Following a similar trend, CPU-accelerated latency experiments on a Intel Core i5-7600K showed that while CAE-MCS achieved an on-chip runtime latency of 7.7$\mu$s, the fastest and slowest OutlierNets in comparison achieved an on-chip latency of 0.366$\mu$s \textbf{(21$\times$ faster)} and 0.746$\mu$s \textbf{ ($\sim$10.3$\times$ faster)}, respectively.

Third, this difference in inference latency holds even when the models are deployed on an ARM Cortex A-72, a significantly slower and power constrained device when compared to the desktop class Intel Core i5. Specifically, we note that the CAE-MCS model now has a on-device latency of 78 $\mu$s while the fastest and slowest OutlierNets achieve 7.2 $\mu$s (\textbf{10.8 $\times$} faster) and 19.1 $\mu$s (\textbf{$\sim$4.1 $\times$} faster). Fourth, the OutlierNet architecture designs exhibit much smaller deployment model sizes than CAE-MCS. Specifically, the CAE-MCS architecture translates to a model size of $\sim$15 MB, which may be impractical to deploy for certain on-device edge scenarios, particularly those leveraging microcontrollers. In contrast, the smallest OutlierNet (at \textbf{2.7 KB}) can fit within the Static RAM of most microcontrollers whereas the larger OutlierNets are within reach of most higher-end microcontrollers such as the STM32F7 series. The low memory requirements and the demonstrated microsecond level on-chip latency on embedded hardware results in systems that can be deployed on factory floors for automated AAD.
\vspace{-0.05in}
\section{Conclusion}
\vspace{-0.05in}
We explored a machine-driven design exploration approach that leverages both human experience and knowledge with the speed and meticulousness of a machine to produce highly compact deep convolutional autoencoder architectures tailored for the purpose of AAD on embedded devices. The resulting OutlierNets possess parameter counts ranging from 686 to 69,877, which translate to model sizes of 2.7 KB to 273 KB. With such low resource requirements, these models can fit within the SRAM of many microcontrollers available today. Despite extremely low complexities, OutlierNets match or exceed the AUC of a much larger convolutional autoencoder architecture while exhibiting microsecond scale latency on embedded hardware. As future work, we aim to study the types of architectures that can be created using machine-driven design exploration with other machine operating sound types, and study design exploration considerations tailored for even lower-power microcontrollers to facilitate compact and real-time AAD on edge devices.

\bibliographystyle{IEEEtran}
\bibliography{references}

\begin{thebibliography}{10}
\providecommand{\url}[1]{#1}
\csname url@samestyle\endcsname
\providecommand{\newblock}{\relax}
\providecommand{\bibinfo}[2]{#2}
\providecommand{\BIBentrySTDinterwordspacing}{\spaceskip=0pt\relax}
\providecommand{\BIBentryALTinterwordstretchfactor}{4}
\providecommand{\BIBentryALTinterwordspacing}{\spaceskip=\fontdimen2\font plus
\BIBentryALTinterwordstretchfactor\fontdimen3\font minus
  \fontdimen4\font\relax}
\providecommand{\BIBforeignlanguage}[2]{{%
\expandafter\ifx\csname l@#1\endcsname\relax
\typeout{** WARNING: IEEEtran.bst: No hyphenation pattern has been}%
\typeout{** loaded for the language `#1'. Using the pattern for}%
\typeout{** the default language instead.}%
\else
\language=\csname l@#1\endcsname
\fi
#2}}
\providecommand{\BIBdecl}{\relax}
\BIBdecl

\bibitem{Purohit2019MIMIIInspection}
\BIBentryALTinterwordspacing
H.~Purohit, R.~Tanabe, K.~Ichige, T.~Endo, Y.~Nikaido, K.~Suefusa, and
  Y.~Kawaguchi, ``{MIMII Dataset: Sound Dataset for Malfunctioning Industrial
  Machine Investigation and Inspection},'' \emph{arXiv}, 9 2019. [Online].
  Available: \url{http://arxiv.org/abs/1909.09347}
\BIBentrySTDinterwordspacing

\bibitem{Ribeiro2020DeepSounds}
\BIBentryALTinterwordspacing
A.~Ribeiro, L.~M. Matos, P.~J. Pereira, E.~C. Nunes, A.~L. Ferreira, P.~Cortez,
  and A.~Pilastri, ``{Deep Dense and Convolutional Autoencoders for
  Unsupervised Anomaly Detection in Machine Condition Sounds},'' \emph{arXiv},
  6 2020. [Online]. Available: \url{http://arxiv.org/abs/2006.10417}
\BIBentrySTDinterwordspacing

\bibitem{Muller2020AcousticLearning}
\BIBentryALTinterwordspacing
R.~M{\"{u}}ller, F.~Ritz, S.~Illium, and C.~Linnhoff-Popien, ``{Acoustic
  Anomaly Detection for Machine Sounds based on Image Transfer Learning},''
  \emph{arXiv}, 6 2020. [Online]. Available:
  \url{http://arxiv.org/abs/2006.03429
  http://dx.doi.org/10.5220/0010185800490056}
\BIBentrySTDinterwordspacing

\bibitem{Koizumi2019ToyADMOS:Detection}
Y.~Koizumi, S.~Saito, H.~Uematsu, N.~Harada, and K.~Imoto, ``{ToyADMOS: A
  dataset of miniature-machine operating sounds for anomalous sound
  detection},'' in \emph{IEEE Workshop on Applications of Signal Processing to
  Audio and Acoustics}, vol. 2019-Octob.\hskip 1em plus 0.5em minus 0.4em\relax
  Institute of Electrical and Electronics Engineers Inc., 10 2019, pp.
  313--317.

\bibitem{Duman2020AcousticProcesses}
\BIBentryALTinterwordspacing
T.~B. Duman, B.~Bayram, and G.~İnce, ``{Acoustic Anomaly Detection Using
  Convolutional Autoencoders in Industrial Processes},'' in \emph{Advances in
  Intelligent Systems and Computing}, vol. 950.\hskip 1em plus 0.5em minus
  0.4em\relax Springer Verlag, 2020, pp. 432--442. [Online]. Available:
  \url{http://link.springer.com/10.1007/978-3-030-20055-8_41}
\BIBentrySTDinterwordspacing

\bibitem{Bayram2021RealAutoencoders}
\BIBentryALTinterwordspacing
B.~Bayram, T.~B. Duman, and G.~Ince, ``{Real time detection of acoustic
  anomalies in industrial processes using sequential autoencoders},''
  \emph{Expert Systems}, vol.~38, no.~1, p. e12564, 1 2021. [Online].
  Available: \url{https://onlinelibrary.wiley.com/doi/10.1111/exsy.12564}
\BIBentrySTDinterwordspacing

\bibitem{Meire2019ComparisonAssets}
M.~Meire and P.~Karsmakers, ``{Comparison of Deep Autoencoder Architectures for
  Real-time Acoustic Based Anomaly Detection in Assets},'' in \emph{Proceedings
  of the 2019 10th IEEE International Conference on Intelligent Data
  Acquisition and Advanced Computing Systems: Technology and Applications,
  IDAACS 2019}, vol.~2.\hskip 1em plus 0.5em minus 0.4em\relax Institute of
  Electrical and Electronics Engineers Inc., 9 2019, pp. 786--790.

\bibitem{He2016}
\BIBentryALTinterwordspacing
K.~He, X.~Zhang, S.~Ren, and J.~Sun, ``{Deep residual learning for image
  recognition},'' in \emph{Proceedings of the IEEE Computer Society Conference
  on Computer Vision and Pattern Recognition}, vol. 2016-Decem.\hskip 1em plus
  0.5em minus 0.4em\relax IEEE Computer Society, 2016, pp. 770--778. [Online].
  Available: \url{http://image-net.org/challenges/LSVRC/2015/}
\BIBentrySTDinterwordspacing

\bibitem{krizhevsky2012imagenet}
A.~Krizhevsky, I.~Sutskever, and G.~E. Hinton, ``Imagenet classification with
  deep convolutional neural networks,'' \emph{Advances in neural information
  processing systems}, vol.~25, pp. 1097--1105, 2012.

\bibitem{Banbury2021Micronets:Microcontrollers}
C.~Banbury, C.~Zhou, I.~Fedorov, R.~M. Navarro, U.~Thakker, D.~Gope, V.~J.
  Reddi, M.~Mattina, and P.~N. Whatmough, ``Micronets: Neural network
  architectures for deploying tinyml applications on commodity
  microcontrollers,'' 2021.

\bibitem{liu2019darts}
H.~Liu, K.~Simonyan, and Y.~Yang, ``Darts: Differentiable architecture
  search,'' 2019.

\bibitem{Tan2018MnasNet:Mobile}
\BIBentryALTinterwordspacing
M.~Tan, B.~Chen, R.~Pang, V.~Vasudevan, M.~Sandler, A.~Howard, and Q.~V. Le,
  ``{MnasNet: Platform-Aware Neural Architecture Search for Mobile},''
  \emph{Proceedings of the IEEE Computer Society Conference on Computer Vision
  and Pattern Recognition}, vol. 2019-June, pp. 2815--2823, 7 2018. [Online].
  Available: \url{http://arxiv.org/abs/1807.11626}
\BIBentrySTDinterwordspacing

\bibitem{Hsu2018MONAS:Learning}
\BIBentryALTinterwordspacing
C.-H. Hsu, S.-H. Chang, J.-H. Liang, H.-P. Chou, C.-H. Liu, S.-C. Chang, J.-Y.
  Pan, Y.-T. Chen, W.~Wei, and D.-C. Juan, ``{MONAS: Multi-Objective Neural
  Architecture Search using Reinforcement Learning},'' 6 2018. [Online].
  Available: \url{http://arxiv.org/abs/1806.10332}
\BIBentrySTDinterwordspacing

\bibitem{elsken2019efficient}
T.~Elsken, J.~H. Metzen, and F.~Hutter, ``Efficient multi-objective neural
  architecture search via lamarckian evolution,'' 2019.

\bibitem{Wong2018FermiNets:Synthesis}
\BIBentryALTinterwordspacing
A.~Wong, M.~J. Shafiee, B.~Chwyl, and F.~Li, ``{FermiNets: Learning generative
  machines to generate efficient neural networks via generative synthesis},'' 9
  2018. [Online]. Available: \url{https://arxiv.org/abs/1809.05989v2}
\BIBentrySTDinterwordspacing

\bibitem{wong2019netscore}
\BIBentryALTinterwordspacing
A.~Wong, ``Netscore: Towards universal metrics for large-scale performance
  analysis of deep neural networks for practical usage,'' \emph{CoRR}, vol.
  abs/1806.05512, 2018. [Online]. Available:
  \url{http://arxiv.org/abs/1806.05512}
\BIBentrySTDinterwordspacing

\end{thebibliography}

\end{document}